\documentclass{article}

\usepackage[nonatbib,preprint]{neurips_data_2024}
\usepackage{graphicx}
\usepackage{amsmath}

\usepackage[utf8]{inputenc} 
\usepackage[T1]{fontenc}    
\usepackage{hyperref}       
\usepackage{url}            
\usepackage{booktabs}       
\usepackage{amsfonts}       
\usepackage{nicefrac}       
\usepackage{microtype}      
\usepackage{xcolor}         
\usepackage{xspace}
\usepackage{underscore}
\usepackage{biblatex}
\usepackage{listings}

\lstdefinestyle{sqlstyle}{
    language=SQL,
    basicstyle=\ttfamily\small,
    keywordstyle=\color{blue},
    commentstyle=\color{gray},
    tabsize=4,
    breaklines=true
}

\newcommand{\ngals}{286,401\xspace}
\newcommand{\ntrain}{204,573\xspace}
\newcommand{\nval}{40,914\xspace}
\newcommand{\ntest}{40,914\xspace}

\title{GalaxiesML: a dataset of galaxy images, photometry, redshifts, and structural parameters for machine learning}

\author{%
  Tuan Do\\
  Physics and Astronomy Department\\
  UCLA\\
  Los Angeles, CA 90095\\
  \texttt{tdo@astro.ucla.edu} \\
  \And
  Bernie Boscoe \\
  Computer Science Department \\
  Southern Oregon University \\
  Ashland, OR 97520 \\
  \texttt{boscoeb@sou.edu} \\
  \AND
  Evan Jones  \\
  Physics and Astronomy Department\\
  UCLA\\
  Los Angeles, CA 90095\\
  \texttt{ejones@astro.ucla.edu} \\
  \AND
  Yun Qi Li  \\
  Physics and Astronomy Department\\
  UCLA\\
  Los Angeles, CA 90095\\
  \texttt{yunqil@g.ucla.edus} \\  
  \AND
  Kevin Alfaro \\
  Physics and Astronomy Department\\
  UCLA\\
  Los Angeles, CA 90095\\
  \texttt{keal1885@gmail.com} \\    
}

\begin{document}

\maketitle

\begin{abstract}
We present a dataset built for machine learning applications consisting of galaxy photometry, images, spectroscopic redshifts, and structural properties. This dataset comprises 286,401 galaxy images and photometry from the Hyper-Suprime-Cam Survey PDR2 in five imaging filters ($g,r,i,z,y$) with spectroscopically confirmed redshifts as ground truth. Such a dataset is important for machine learning applications because it is uniform, consistent, and has minimal outliers but still contains a realistic range of signal-to-noise ratios. We make this dataset public to help spur development of machine learning methods for the next generation of surveys such as Euclid and LSST.  
The aim of GalaxiesML is to provide a robust dataset that can be used not only for astrophysics but also for machine learning, where image properties cannot be validated by the human eye and are instead governed by physical laws. We describe the challenges associated with putting together a dataset from publicly available archives, including outlier rejection, duplication, establishing ground truths, and sample selection.  This is one of the largest public machine learning-ready training sets of its kind with redshifts ranging from 0.01 to 4. The redshift distribution of this sample peaks at redshift of 1.5 and falls off rapidly beyond redshift 2.5. We also include an example application of this dataset for redshift estimation, demonstrating that using images for redshift estimation produces more accurate results compared to using photometry alone. For example, the bias in redshift estimate is a factor of 10 lower when using images between redshift of 0.1 to 1.25 compared to photometry alone. Results from dataset such as this will help inform us on how to best make use of data from the next generation of galaxy surveys.

\end{abstract}

\begin{figure}[!htb]
\centering
\includegraphics[width=5in]{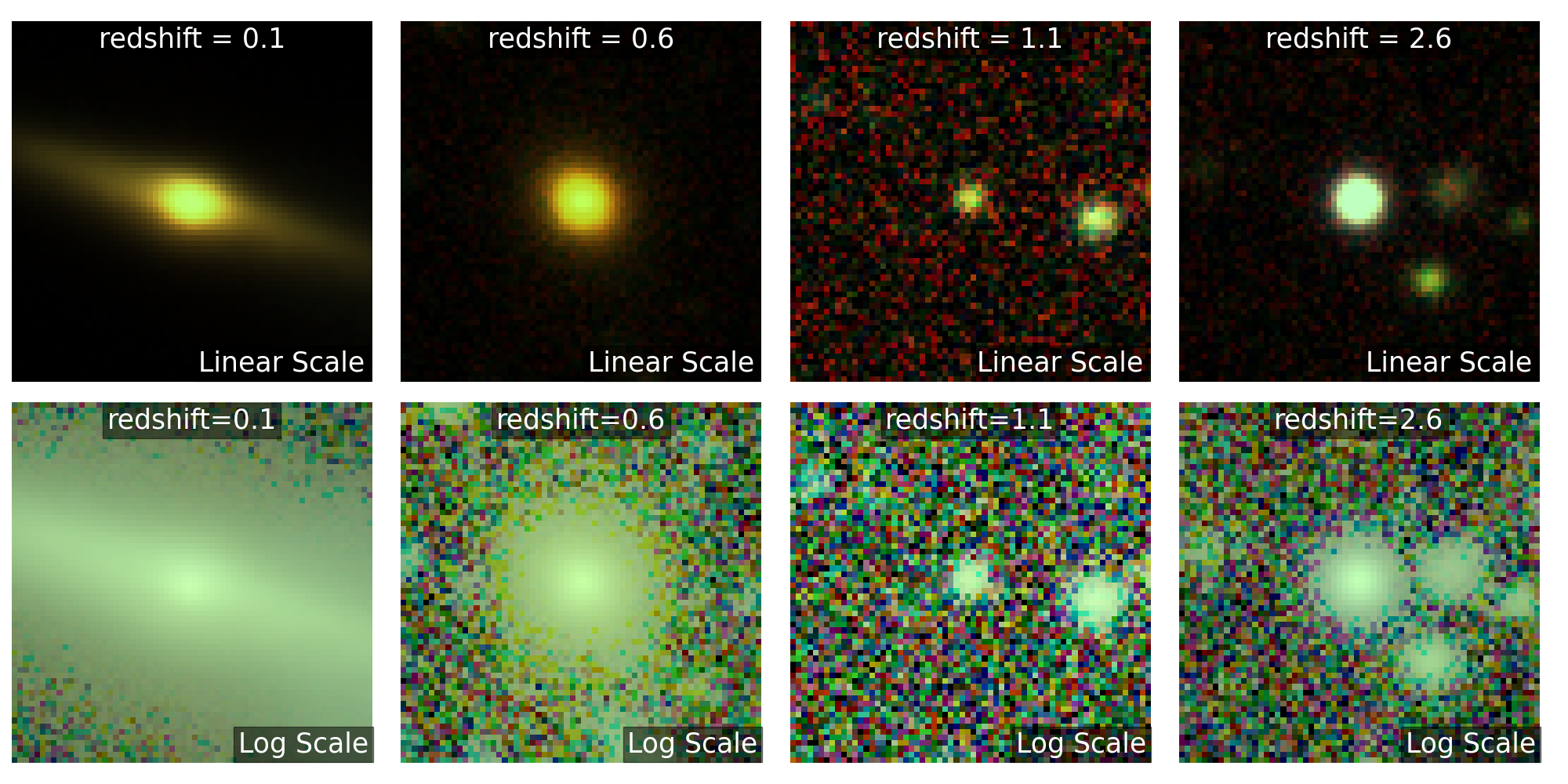}
\caption{Example of galaxies at different redshifts from the GalaxiesML dataset. The top row shows the images in a linear intensity scale while the bottom row shows the images in a logarithmic scale to show lower surface brightness features like nearby galaxies . }
\label{fig:example}
\end{figure}

\section{Introduction}

One of the major questions in physics and astrophysics is the nature of dark matter and dark energy. Together, dark matter and dark energy make up over 95\% of the energy density of the universe, yet we do not know their particle or field nature \cite{planckcollaboration2020}. The most promising approach to investigating their nature is through observations of the Universe on cosmic scales, pushing the limits of not only our telescopes and instrumentation, but also of our data analysis techniques. 

Machine learning holds potential to help achieve the ambitious science goals of the large experiments in Astrophysics that are coming online now and in the next few years. The vast majority of data we have about the Universe is in the form of images, which are not always easily interpreted. For example, a key measurement required is how galaxies are spatially distributed across different cosmic times, but the distance to a galaxy (redshift) is not easily determined from its images alone. Fortunately, because the properties of galaxies evolve over time according to physical laws, the images of galaxies at different wavelengths encode information about how long ago their light was emitted, which in turn enables the determination of their distance. The relationship between the images of galaxies and their redshifts is complex due to the diversity of their formation pathways. Mapping this relationship is an ideal problem for machine learning \cite[e.g.,][]{newman2022a}. 

While astronomy has often led the way in the sciences for open access to data, there are not many datasets specifically designed for machine learning. Astronomical data is typically available through data archives as raw or processed data products for individual objects. Although these data products usually meet the needs of most researchers, machine learning applications require a higher level of data curation and compilation to ensure accurate predictions \cite{boscoe2022}. Creating valid, replicable datasets is often one of the most time-consuming facets of the machine learning process \cite{terrizzano}. 

Large science surveys such as the Large Survey in Space and Time (LSST) \cite{ivezic2008} and the Euclid mission \cite{euclidcollaboration2024} aim to observe billions of galaxies to map their distribution throughout cosmic time, with the goal of constraining models of dark matter and dark energy. These surveys are expected to produce data at scales orders of magnitude larger than what we currently possess.

To efficiently analyze and leverage these vast datasets, astronomers have increasingly turned to machine learning methods. Machine learning models are particularly useful for problems where calculating likelihoods is challenging, especially when large datasets for training and sufficient computing resources are available. Extracting information from images falls into this category. For instance, accurately estimating redshifts from images is critical for achieving LSST's cosmology goals. spectroscopy is too time-consuming and expensive to apply to the billions of galaxies required for measuring sensitive cosmological parameters \cite[e.g.,][]{mandelbaum2018e}. Because it is unclear how to analytically compute redshifts from vast numbers of images, machine learning offers a data-driven solution to this challenge. In astronomy, images are a primary source of information, but their resolutions vary as more distant objects are captured.  As the data collection grows, the task of ground truthing continually evolves. Human judgment cannot be relied upon to assess images 'by eye' in traditional machine learning benchmarking methods; instead, other methods must be used to determine the accuracy of models' predictions.

In this work, we present GalaxiesML, a new publicly available dataset of \ngals galaxy images in five wavelength filters and galaxy properties specifically designed for machine learning applications. While the dataset has primarily been used for photometric redshift estimation, it is versatile and can support other science goals as well. It is based on publicly available survey data, including images, photometry, and spectroscopic redshifts. The spectroscopic redshifts serve as the ground truth for the redshifts associated with the images. Additional processing has been performed to measure the morphological information for each galaxy image. We chose to use the Hyper-Suprime-Cam (HSC) Survey \cite{aihara2019} as the basis for this dataset because it provides a representative sample of galaxies comparable to those expected from the next generation of large sky surveys. GalaxiesML is optimized for machine learning models by carefully addressing issues like outliers and missing data, and the images and tabular data are provided in a format that can be easily integrated into modern machine learning frameworks.

By releasing this dataset, we aim to provide: (1) a large source of training data, (2) a consistent dataset for model comparisons, (3) a reduction in barriers to entry for applying machine learning in cosmology, and (4) a fixed dataset that enables reproducibility of findings. We also hope that this dataset will be valuable to general machine learning practitioners as an alternative to other imaging datasets, offering data from a scientific domain where the images are directly related to physical laws.

\section{Related Work}
Astronomers have been using machine learning for decades. Examples from the past decade include the morphological classification of galaxies using training data crowd-sourced from GalaxyZoo \cite{lintott2008} which demonstrated that deep neural networks can perform as accurately as humans in classifying galaxies \cite{dieleman2015a}. In exoplanet science, \cite{shallue2018a} showed that convolutional neural networks can effectively detect planets in light curves measured by the Kepler spacecraft.
The AstroML project \cite{ivezic2014} has compiled both analysis tools and datasets that are useful for machine learning in astrophysics.
However, these datasets are relatively small and are primarily intended for demonstrating how models work. They do not present significant challenges for new deep learning models nor do they represent the realistic test cases that will arise from the next generation of telescopes. Additionally, the datasets used in machine learning research papers in astronomy are often not published or are specifically designed to test a single algorithm \cite{beck2021,bilicki2018,bonnett2015a,cavuoti2012,disanto2018a,eriksen2020,henghes2022,mu2020,pasquet2019,schuldt2021a,stivaktakis2020,tanaka2018g,wright2019}. Despite major advances in available quality data and sophisticated tooling, the growth of publicly available astronomy machine learning datasets has been slow. For example, there are no astronomy datasets available in the Tensorflow dataset repository or the University of California, Irvine (UCI) machine learning repository. While Kaggle hosts several toy astronomy datasets in various user repositories, none of them are comparable to GalaxiesML in terms of scale and scientific applicability.

\section{The GalaxiesML Dataset}

\label{sec:data}

The primary data sources for this work are from the HSC Survey Data Release 2 and the associated spectroscopic redshift database\cite{aihara2019}. This survey is conducted using the 8-meter diameter Subaru Telescope, located on Maunakea, a mountain in Hawaii. 
The HSC survey PDR2 is a sky survey over 300 square degrees of the sky and contains images of over 30 million galaxies in the $g,r,i,z,y$ imaging filters. The HSC survey reaches a similar depth as upcoming large sky surveys, but over smaller region of the sky, which makes this survey a good precursor to train and test machine learning models. The HSC PDR2 database contains spectroscopic redshifts cross-matched by the team (within a projected distance of $< 0.5^{\prime\prime}$)  to the HSC catalog using of publicly available spectroscopic redshift catalogs \cite{lilly2009,bradshaw2013,mclure2012,skelton2014,momcheva2016,lefevre2013,garilli2014,liske2015,davis2003,newman2013,coil2011,cool2013}.

\begin{figure}[!htb]
 \center
 \includegraphics[width=3.25in]{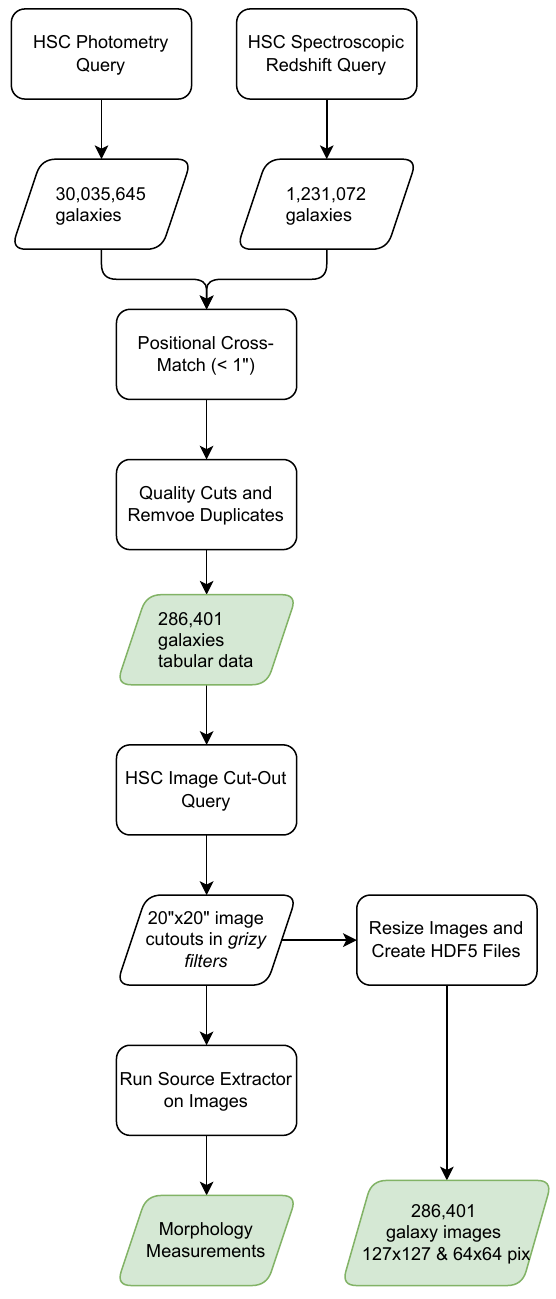}
 \caption{Flow chart showing the steps used in creating the GalaxiesML dataset. Rectangles represent processes and parallelograms are the products. The green parallelograms are the datasets that are part of the release.}
 \label{fig:flowchart}
\end{figure}

We assemble the dataset in 6 major stages: 
\begin{enumerate}
    \item Query and download from the HSC PDR2 and spectroscopic redshift databases
    \item Apply additional data quality filters \& remove duplicates and outliers
    \item Download images and produce cutouts
    \item Fit images to determine morphological information
    \item Save the dataset into ML compatible formats
\end{enumerate}
These stages are shown in the flow chart in Fig. \ref{fig:flowchart} and outlined below.

\subsection{Database Queries}
We create a custom SQL query to select and download data from the HSC Survey PDR2 Archive \cite{aihara2019}. The initial selection of galaxies is designed to include as many well-observed galaxies as possible. We use the following criteria for selecting galaxies from the PDR2 database:
\begin{itemize}
\item \texttt{grizy\_cmodel\_flux\_flag = False} $z > 0$
\item \texttt{grizy\_pixelflags\_edge} = False  
\item \texttt{grizy\_pixelflags\_interpolatedcenter} = False 
\item \texttt{grizy\_pixelflags\_saturatedcenter} = False, \text{unique galaxy object ID}
\item \texttt{grizy\_pixelflags\_crcenter} = False, 
\item \texttt{grizy\_pixelflags\_bad} = False
\item \texttt{grizy\_sdsscentroid\_flag} = False
\end{itemize}
For photometric redshift applications, the dataset needs a source of 'truth' for the redshift of each galaxy, so we also require that the objects have reliable spectroscopic redshift measurements by joining the tables with the spectroscopic redshift tables in the HSC database. We use the spectroscopic redshift information gathered by the HSC team as the ground truth for this sample. For detail of this process, see \cite{aihara2019}. In brief, a match was made between the location of objects in the HSC survey with those from multiple spectroscopic surveys. We find that sample of objects from step X has Y counterparts with spectroscopic redshift information. In addition to joining with the photometry table, we also apply the following filters for the redshifts:
\begin{itemize}
\item $z > 0$
\item $z \ne 9.9999$
\item $0 < z_{err} < 1$
\item \texttt{specz\_flag\_homogeneous = True}
\end{itemize}

We require that the galaxies be detected in all 5 imaging filters. The database query is reproduced in Appendix A. Overall, this initial sample has 801,246 objects. 

\subsection{Additional data quality filters \& remove duplicates}

We applying the following additional filters to the list of galaxies after extract the objects from the database:

\begin{itemize}
    \item Redshift range: $ 0.01 < z < 4.0$
    \item Spectroscopic redshift error: $\sigma_{specz} < 0.005/(1+specz)$
    \item Magnitude range in all bands: $0 <$ \texttt{\{band\}\_cmodel \_mag} $< 50$
\end{itemize}

To build our final sample, we then remove duplicate objects from the sample. While each row has a unique ObjID in the database, there can be multiple Object IDs for the same physical source because there were multiple measurements of its photometry. About 70\% of the entries do not refer to unique sources. We define duplicates as objects that have the same spectroscopic redshift identifier. Note that the spectroscopic redshifts were matched to the photometry using a distance of 0.5 arcseconds by the HSC team \footnote{\url{https://hsc-release.mtk.nao.ac.jp/doc/index.php/dr1_specz/}}. We also identify duplicates sources with the same HSC object ID, but different spectroscopic IDs. In the case of duplicates, we keep the first match and remove the others. After this stage, our final sample includes \ngals sources.

\subsection{Download and produce image cutouts}

After obtaining the final sample of galaxies, we query the HSC PDR2 cutout service to download the images\footnote{\url{https://hsc-release.mtk.nao.ac.jp/das_cutout/pdr2/}} (see Supplemental). We submit queries at the RA and DEC for each band in batches of 100,000 galaxies at a time, with cutout sizes of $10^{\prime\prime}\times10^{\prime\prime}$. We download the \texttt{coadd} option for images and selected the \texttt{PDR2 Wide} option. The images are downloaded as FITS files, with one in each band.

\begin{figure}[tbh]
 \center
 \includegraphics[width=5in]{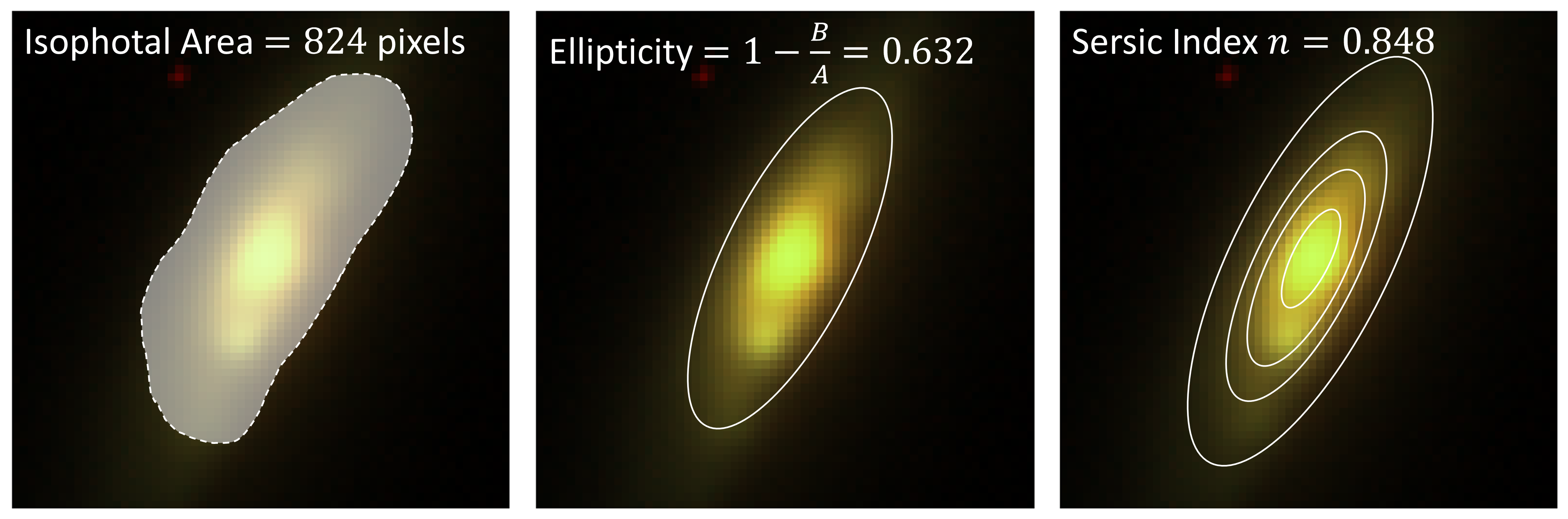}
 \caption{Example of the morphological parameters measured on a low redshift galaxy (Object ID 36416246018753893, $z = 0.0713$) using Source Extractor. \textbf{Left:} isophototal area, \textbf{center:} ellipticity, \textbf{right:} Sersic Index.}
 \label{fig:morphology}
\end{figure}

\subsection{Measurement of the Morphological Parameters}

We also extracted typical morphological features from the galaxy images to aid in interpreting the images and models. The 127$\times$127 pixel images were fit using Source Extractor \cite{bertin1996}. Source Extractor is a tool that is often used to model galaxies in images. Source Extractor fits the pixel values of images using parameterized models of galaxies using an estimate of the point spread function (PSF). It can fit for multiple sources at once and produce a segmentation map that indicates which pixels belong to which source. We parameterized the morphology of the galaxies using several models:
\begin{itemize}
    \item Elliptical model - fitting the sources as an ellipse with a semi-major axis, semi-minor axis, and orientation. 
    \item Sersic model - fit the flux distribution with a Sersic profile
    \item Isophotal, half-light, and Petrosian radius.
\end{itemize}

We also determine the number of galaxies in the image from the image segmentation map from Source Extractor. Source Extractor identifies sources using a detection threshold of pixel values above the background (called \texttt{DETECT\_THRESH}), which we set to 3$\sigma$. We use the source position parameters \texttt{X\_IMAGE} and \texttt{Y\_IMAGE} closest to the center of the image as the galaxy that is associated with the spectroscopic redshift from the HSC catalog. We also utilize the position parameters to identify the number of other galaxies in a circle with a radius of 15, 10, and 5 pixels around the center of the image to quantify the number of nearby sources. 
The complete Source Extract configuration file we use for each galaxy is reproduced in Supplemental Materials. 

\subsection{Save the dataset into ML compatible format}

The imaging data is stored in the HDF5 file format. This file contains cutouts of each galaxy image in $g, r, i, z, y$, stored in the \texttt{image} key of the HDF5 file as an $N_{gal}\times N_x \times N_y$ array. The decision to use HDF5 is due to its support for reading in only parts of the dataset at a time, and for its ease of use in machine learning frameworks. To create the HDF5, we download the images of each object in all five HSC imaging filters in the FITS format and combine the data into HDF5 files. The data are downloaded as cutouts from larger scale HSC images using the HSC cutout service \footnote{\url{https://hsc-release.mtk.nao.ac.jp/das_cutout/pdr2/}}. We use an image radius query of \texttt{sh = 10arcseconds, sw = 10arcseconds}, which results in images of $20 \times 20$ arcseconds in spatial dimension. The images have a plate scale of 0.168 arseconds per pixel \cite{aihara2018}. We created two image sizes: 127x127 pix and 64x64 pix. Most machine learning methods require all images to have fixed sizes. Having multiple options for the image sizes allows one to test the effect of image sizes on model performance and to use wider variety of models. For each image size we create 3 HDF5 files for training (60\%), validation (20\%), and testing (20\%) by randomly splitting the data. We perform the split for convenience and to more easily compare model performances with the same split. 

\section{Description of ML Dataset}
\label{sec:description}

The GalaxiesML dataset is a collection of tabular data, imaging data, and metadata. The tabular data is organized as a CSV file and also included in the HDF5 files with the images. In the HDF5, the images ($5\times127\times127$ or $5\times64\times64$) are under the \texttt{image} key, while the other tabular data are under the keys corresponding to their column name. The tabular data providing detailed information on the identification and characteristics of each galaxy sourced from the HSC and spectroscopic database. The tables also include the extracted features including morphology. 

\begin{table}[]
\caption{GalaxiesML Column Definition - Galaxy Properties \& Morphology Measurements}
\begin{tabular}{lcl}
\hline\hline
Column Name & Units & Description \\ \hline 
\texttt{object_id} & &  object ID from the HSC survey. Unique ID in 64bit integer \\
\texttt{coord} & (deg, deg, deg) & Coordinate used in coneSearch(coord, RA, DEC, RADIUS) \\
\texttt{ra} & deg & RA (J2000.0) of the image center \\
\texttt{dec} & deg & DEC (J2000.0) of the image center \\
\texttt{\{band\}_cmodel_mag} & mag & magnitude of the central galaxy in filter \{\texttt{band}\} \\
\texttt{\{band\}_cmodel_magsigma} & mag & uncertainty in the magnitude in filter \{\texttt{band}\} \\
\texttt{skymap_id} & & location of the galaxy in internal survey position definition (tract, patch) \\
\texttt{specz_name} & & name(s) of the galaxy in the spectroscopic survey(s) \\
\texttt{specz_flag_homogeneous} & & Homogenized spec-z flag. (TRUE=secure, FALSE=insecure) \\
\texttt{specz_mag_i} & mag &  i-band magnitude of the galaxy in the spectroscopic survey \\
\texttt{specz_ra} & deg & RA (J2000.0) of galaxy in spectroscopic survey  \\
\texttt{specz_dec} & deg & DEC (J2000.0) of galaxy in spectroscopic survey \\
\texttt{specz_redshift} & & spectroscopic redshift \\
\texttt{specz_redshift_err} & & spectroscopic redshift uncertainty \\
\texttt{\{band\}_central_image_pol_15px_rad} & &  See Section \ref{sec:description} \\
\texttt{\{band\}_central_image_pop_10px_rad} & &   See Section \ref{sec:description} \\
\texttt{\{band\}_central_image_pop_5px_rad} & &  See Section \ref{sec:description} \\
\texttt{\{band\}_ellipticity} & pixels &  See Section \ref{sec:description} \\
\texttt{\{band\}_half_light_radius} & pixels & See Section \ref{sec:description} \\
\texttt{\{band\}_isophotal_area} & pixels & See Section \ref{sec:description} \\
\texttt{\{band\}_major_axis} & pixels & See Section \ref{sec:description} \\
\texttt{\{band\}_minor_axis} & pixels & See Section \ref{sec:description} \\
\texttt{\{band\}_peak_surface_brightness} & mag/sq. arcsec & See Section \ref{sec:description} \\
\texttt{\{band\}_petro_rad} & pixels & See Section \ref{sec:description} \\
\texttt{\{band\}_pos_angle} & deg & See Section \ref{sec:description} \\
\texttt{\{band\}_sersic_index} & & See Section \ref{sec:description} \\
\texttt{\{band\}_total_galaxies} & & See Section \ref{sec:description} \\ \hline
\end{tabular}
\label{tab:galaxyproperties}
\end{table}

The following columns are the list of extracted parameters using Source Extractor:

\begin{itemize}

\item \texttt{\{band\}_central_image_pol_15px_rad}: The number of detected objects within a 15-pixel-radius circle, centered at the middle of the image. Derived from Source Extractor segmentation file.

\item \texttt{\{band\}_central_image_pop_10px_rad}: The number of detected objects within a 10-pixel-radius circle, centered at the middle of the image. Derived from Source Extractor segmentation file.

\item \texttt{\{band\}_central_image_pop_5px_rad}: The number of detected objects within a 5-pixel-radius circle, centered at the middle of the image. Derived from Source Extractor segmentation file.

\item \texttt{\{band\}_ellipticity}: The ellipticity of the object, defined as 1 - B/A. Where B is the semi-minor axis of an object and A is the semi-major axi (pixels).

\item \texttt{\{band\}_half_light_radius}: The radius of an object at which 50\% of the flux is contained (pixels).

\item \texttt{\{band\}_isophotal_area}: The total number of pixels that a detected object is composed of.

\item \texttt{\{band\}_major_axis}: Major axis of the detected object (pixels).

\item \texttt{\{band\}_minor_axis}: Minor axis of the detected object (pixels).

\item \texttt{\{band\}_peak_surface_brightness}: The peak surface brightness above background of the object (magnitudes per square arcsecond).

\item \texttt{\{band\}_petro_rad}: Petrosian radius of an object (pixels).

\item \texttt{\{band\}_pos_angle}: Rotation of the major axis with respective to the x-axis of the image plane, counterclockwise (degrees).

\item \texttt{\{band\}_sersic_index}: The Sérsic index of the object, which describes the shape of the object's light profile.

\item \texttt{\{band\}_total_galaxies}: Total number of galaxies detected by Source Extractor in an image.

\end{itemize}

The data is available from Zenodo with a DOI: 10.5281/zenodo.11117528. 

\section{GalaxiesML Use: Neural Networks for Redshift Predictions}

\begin{figure}[tbh]
 \center
 \includegraphics[width=\linewidth]{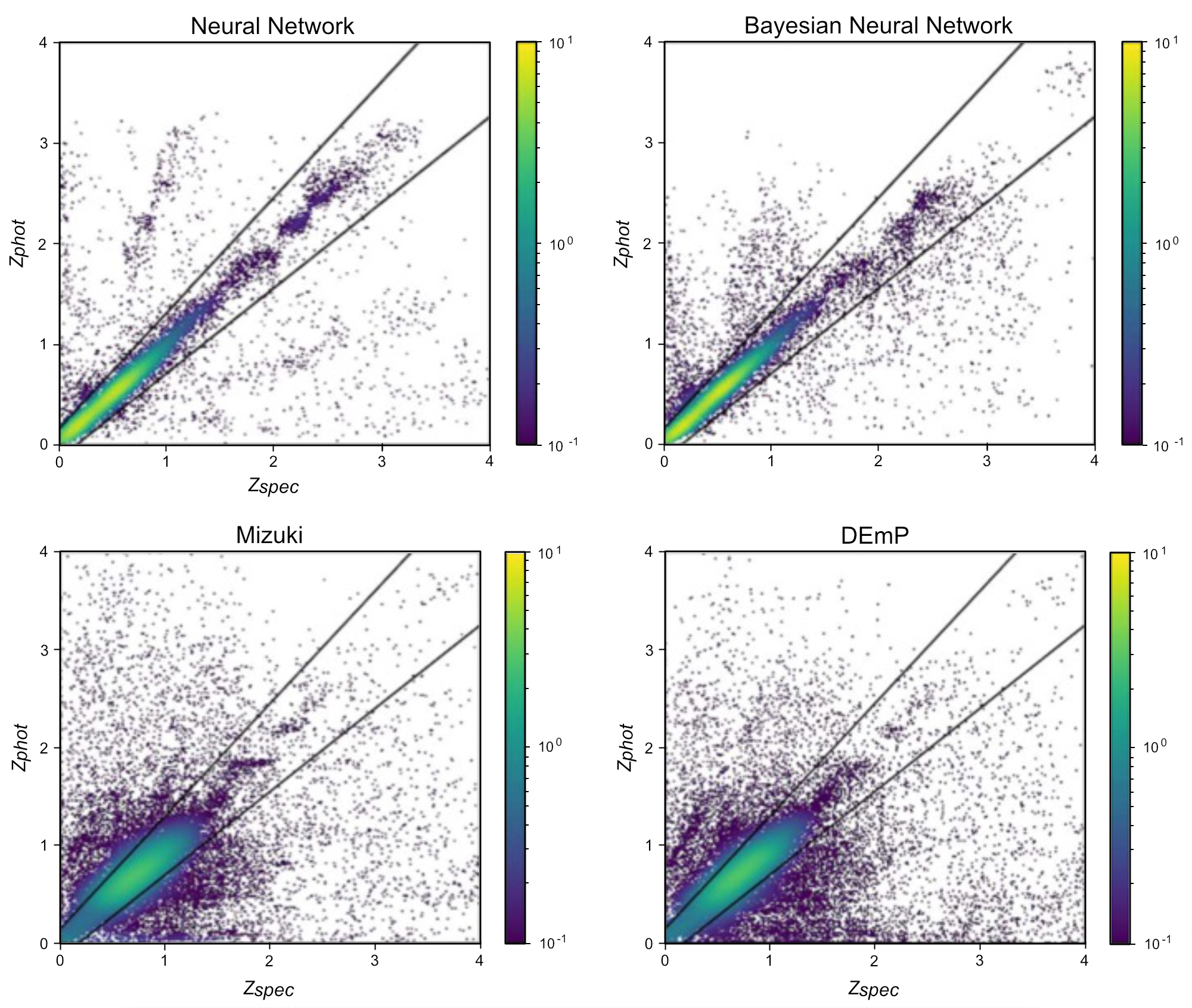}
 \caption{Visualization of predicted photo-zs vs. measured spectroscopic redshifts for a non-probabilistic neural network (top left), a Bayesian neural network (top right), and two previously proposed ML models (bottom left and right). Figure from \cite{jones2022}}.
 \label{fig:jones}
\end{figure}

The publications using GalaxiesML include that of \cite{jones2022}, which developed the first  Bayesian neural networks for photometric redshifts (Fig. \ref{fig:jones}). This work found that Bayesian neural networks can provide accurate uncertainties for the redshift predictions, which is crucial for scientific use of the predictions in cosmology. In addition \cite{jones2024} found these networks were mostly robust to out-of-distribution biases and can provide good uncertainty estimates even for faint and low signal to noise galaxies. 

\section{Discussion and Conclusion}
We have introduced GalaxiesML, a machine learning ready dataset designed for use by both astrophysicists or learners desiring to use a dataset with a scientific aim that includes noise and a ground truth derived from spectroscopic data rather than human visual assessment. GalaxiesML can be used in a variety of applications with or without the inclusion of image data. The dataset's size, $\sim$300,000 galaxies and over 100 features allows for scientific exploration that is "medium sized" (116GB) that can be efficiently loaded into common machine learning tools, such as Jupyter notebooks, or run on GPU- enabled machines with a reasonable number of cores. It is designed to be within the reach of individual researchers while also being publicly accessible.

Our goal is to provide additional resources to make this dataset useful for scientific exploration, education, and benchmarking. We also aim to ensure compatibility with popular dataloader APIs, facilitating easier integration into machine learning workflows. GalaxiesML represents an important contribution to the growing collection of scientific datasets designed specifically for machine learning applications.

\clearpage

\printbibliography

\section{Appendix}

In this Appendix, we present additional details for constructing the GalaxiesML dataset and show an example of its use. 

\section{HSC Database Queries}

To obtain the initial list of galaxies for Step 1, we query the HSC PDR2 database to cross-match data from the HSC PDR2 imaging and photometry dataset with that of the spectroscopic dataset. Specifically, this query uses the \texttt{pdr2_wide.forced} and \texttt{pdr2_wide.specz} tables. The query also includes data quality filters as described in the main paper. 

\label{appendix:sql}
\begin{lstlisting}[style=sqlstyle]
SELECT 
    object_id 
	  , specz_redshift_err
      , specz_redshift
	  , specz_mag_i
      , specz_name
      , specz_ra 
      , specz_dec,
specz_flag_homogeneous,
	  ra,
	  dec,
	  coord,
	  skymap_id,  
	  g_cmodel_mag
	  	  , r_cmodel_mag
		  	  , i_cmodel_mag
			  	  , z_cmodel_mag
				  	  , y_cmodel_mag
					  
	
		  , g_cmodel_magsigma
	  	  , r_cmodel_magsigma
		  	  , i_cmodel_magsigma
			  	  , z_cmodel_magsigma
				  	  , y_cmodel_magsigma


	


    FROM pdr2_wide.forced
	

	
	LEFT JOIN pdr2_wide.forced2 USING (object_id)
	LEFT JOIN pdr2_wide.specz USING (object_id)
	
		WHERE
 	NOT
	g_sdsscentroid_flag
		AND NOT
	r_sdsscentroid_flag
		AND NOT
	i_sdsscentroid_flag
		AND NOT
	z_sdsscentroid_flag
		AND NOT
	y_sdsscentroid_flag
	
	AND NOT
		g_pixelflags_interpolatedcenter
	AND NOT
	r_pixelflags_interpolatedcenter
	AND NOT
	i_pixelflags_interpolatedcenter
	AND NOT
	z_pixelflags_interpolatedcenter
	AND NOT
	y_pixelflags_interpolatedcenter
	AND NOT
		g_pixelflags_saturatedcenter
	AND NOT
	r_pixelflags_saturatedcenter
	AND NOT
	i_pixelflags_saturatedcenter
	AND NOT
	z_pixelflags_saturatedcenter
	AND NOT
	y_pixelflags_saturatedcenter
	AND NOT
    g_cmodel_flag
    AND NOT
    r_cmodel_flag
    AND NOT
    i_cmodel_flag
    AND NOT
    z_cmodel_flag
    AND NOT
    y_cmodel_flag
	AND NOT
	g_pixelflags_edge
	AND NOT
	r_pixelflags_edge	
	AND NOT
	i_pixelflags_edge		
	AND NOT
	z_pixelflags_edge
	AND NOT
	y_pixelflags_edge
	AND NOT 	
	g_pixelflags_crcenter
	AND NOT
	r_pixelflags_crcenter	
	AND NOT
	i_pixelflags_crcenter		
	AND NOT
	z_pixelflags_crcenter
	AND NOT
	y_pixelflags_crcenter
	AND NOT 	
	g_pixelflags_bad
	AND NOT
	r_pixelflags_bad	
	AND NOT
	i_pixelflags_bad		
	AND NOT
	z_pixelflags_bad
	AND NOT
	y_pixelflags_bad
	AND  specz_redshift > 0
	AND 0 < specz_redshift_err
	AND specz_redshift_err < 1
	AND specz_redshift < 9.999

 
\end{lstlisting}

\section{HSC Image Query}
Images were obtained by uploading the ra and dec positions in batches of 100,00 requests at a time to the HSC Image Cutout Service at:
\url{https://hsc-release.mtk.nao.ac.jp/das_cutout/pdr2/manual.html#list-to-upload}. Here is a example of some of the lines from the positional query:
\begin{lstlisting}[basicstyle=\small]
#? rerun	filter	ra	dec	sw	sh
pdr2_wide	HSC-Y	31.73471487	-6.610750394	10arcseconds	10arcseconds
pdr2_wide	HSC-Y	31.19229445	-6.44807734	10arcseconds	10arcseconds
pdr2_wide	HSC-Y	31.20098659	-6.878016245	10arcseconds	10arcseconds
pdr2_wide	HSC-Y	31.09620369	-6.250176133	10arcseconds	10arcseconds
pdr2_wide	HSC-Y	31.16933934	-6.523789799	10arcseconds	10arcseconds
\end{lstlisting}

\section{Source Extractor Configuration}
\label{appendix:sourceextractor}
We use the software Source Extractor \cite{bertin1996} to measure the shape and flux distribution of the galaxies in the GalaxiesML dataset. The fit parameters are described in the main text. Source Extract requires both a parameter file and a configuration file to run. Here we give examples of these files. 

We use the following parameter file running Source Extractor on each image:
\begin{lstlisting}[basicstyle=\small]
NUMBER    1               #Running object number
DETECT_THRESH 3
CATALOG_NAME output_image
CATALOG_TYPE ASCII_HEAD
CHECKIMAGE_TYPE  SEGMENTATION
PETRO_RADIUS 1
PETRO_TYPE AUTO
X_IMAGE 60
Y_IMAGE 60
XMIN_IMAGE 57.5
XMAX_IMAGE 62.5
YMIN_IMAGE 57.5
YMAX_IMAGE 62.5
ISOAREA_IMAGE 0.0
ISOAREA_WORLD 0.0
A 0.0
B 0.0
A_IMAGE
B_IMAGE
THETA_IMAGE 0.0
THETA_WORLD 0.0
MU_MAX 0.0
ELLIPTICITY 0.0
SERSIC 0.0
PHOT_TYPE SERSIC
PHOT_AUTOPARAMS 2.5, 3.5
PHOT_APERTURES 5
FLUX_RADIUS
SPHEROID_SERSICN 4.0
SPHEROID_RE 10.0
\end{lstlisting}

We use the following configuration file running Source Extractor on each image:
\begin{lstlisting}[basicstyle=\small]
# Default configuration file for SExtractor 2.25.0
# EB 2021-05-31
NUMBER    1
PETRO_TYPE AUTO
PETRO_THRESH 0.2
#-------------------------------- Catalog ------------------------------------

CATALOG_NAME     test.cat       # name of the output catalog
CATALOG_TYPE     ASCII_HEAD     # NONE,ASCII,ASCII_HEAD, ASCII_SKYCAT,
                                # ASCII_VOTABLE, FITS_1.0 or FITS_LDAC
PARAMETERS_NAME  default.param  # name of the file containing catalog contents

#------------------------------- Extraction ----------------------------------

DETECT_TYPE      CCD            # CCD (linear) or PHOTO (with gamma correction)
DETECT_MINAREA   5              # min. # of pixels above threshold
DETECT_THRESH    1.5            # <sigmas> or <threshold>,<ZP> in mag.arcsec-2
ANALYSIS_THRESH  1.5            # <sigmas> or <threshold>,<ZP> in mag.arcsec-2

FILTER           Y              # apply filter for detection (Y or N)?
FILTER_NAME      gauss_1.5_3x3.conv   # name of the file containing the filter

DEBLEND_NTHRESH  32             # Number of deblending sub-thresholds
DEBLEND_MINCONT  0.005          # Minimum contrast parameter for deblending

CLEAN            Y              # Clean spurious detections? (Y or N)?
CLEAN_PARAM      1.0            # Cleaning efficiency

MASK_TYPE        CORRECT        # type of detection MASKing: can be one of
                                # NONE, BLANK or CORRECT

#------------------------------ Photometry -----------------------------------

PHOT_APERTURES   5              # MAG_APER aperture diameter(s) in pixels
PHOT_AUTOPARAMS  2.5, 3.5       # MAG_AUTO parameters: <Kron_fact>,<min_radius>
PHOT_PETROPARAMS 2.0, 3.5       # MAG_PETRO parameters: <Petrosian_fact>,
                                # <min_radius>

SATUR_LEVEL      50000.0        # level (in ADUs) at which arises saturation
SATUR_KEY        SATURATE       # keyword for saturation level (in ADUs)

MAG_ZEROPOINT    0.0            # magnitude zero-point
MAG_GAMMA        4.0            # gamma of emulsion (for photographic scans)
GAIN             0.0            # detector gain in e-/ADU
GAIN_KEY         GAIN           # keyword for detector gain in e-/ADU
PIXEL_SCALE      0.17            # size of pixel in arcsec (0=use FITS WCS info)

PHOT_TYPE SERSIC
SERSIC_FIT Y
FIT_PROFILE N
#------------------------- Star/Galaxy Separation ----------------------------

SEEING_FWHM      1.2            # stellar FWHM in arcsec
STARNNW_NAME     default.nnw    # Neural-Network_Weight table filename

#------------------------------ Background -----------------------------------

BACK_SIZE        64             # Background mesh: <size> or <width>,<height>
BACK_FILTERSIZE  3              # Background filter: <size> or <width>,<height>

BACKPHOTO_TYPE   GLOBAL         # can be GLOBAL or LOCAL

#------------------------------ Check Image ----------------------------------

#CHECKIMAGE_TYPE  -BACKGROUND           # can be NONE, BACKGROUND, BACKGROUND_RMS,
                                # MINIBACKGROUND, MINIBACK_RMS, -BACKGROUND,
                                # FILTERED, OBJECTS, -OBJECTS, SEGMENTATION,
                                # or APERTURES
#CHECKIMAGE_NAME  check.fits     # Filename for the check-image

#--------------------- Memory (change with caution!) -------------------------

MEMORY_OBJSTACK  3000           # number of objects in stack
MEMORY_PIXSTACK  300000         # number of pixels in stack
MEMORY_BUFSIZE   1024           # number of lines in buffer

#----------------------------- Miscellaneous ---------------------------------

VERBOSE_TYPE     NORMAL         # can be QUIET, NORMAL or FULL
HEADER_SUFFIX    .head          # Filename extension for additional headers
WRITE_XML        N              # Write XML file (Y/N)?
XML_NAME         sex.xml        # Filename for XML output
DETECT_THRESH 3
NUMBER    1                     #Running object number
DETECT_THRESH 3
CATALOG_NAME output_image
CATALOG_TYPE ASCII_HEAD
CHECKIMAGE_TYPE  SEGMENTATION
ERRXY_IMAGE 5    
\end{lstlisting}

\section{Redshift estimation using convolutional neural network with images and photometry}
\label{sec:redshift}

We demonstrate the improvements that can be made by including imaging information by developing a hybrid convolutional neural network (CNN) that includes both images and photometry. We use a CNN architecture with 7 convolutional layers (with $3\times4$ kernels) and 5 pooling layers, inter-dispersed between the convolutional layers. A novel aspect of our network is the inclusion of photometry as well, which has inputs for the five-band photometry and 6 dense layers. The photometric network is connected to the output of the CNN, which then goes through three additional fully connected layers. 

\begin{figure}[tbh]
 \center
 \includegraphics[width=3.75in]{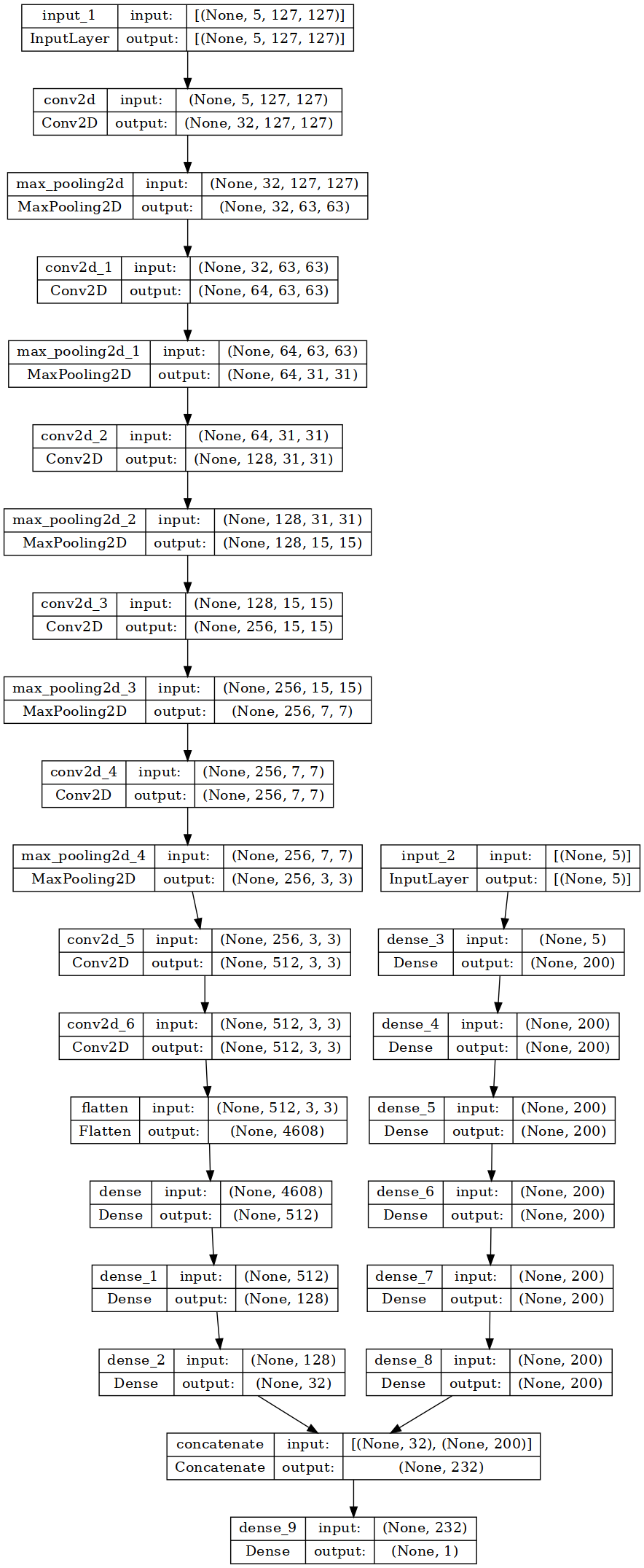}
 \caption{CNN architecture for this work. The model includes both images and photometry.}
 \label{fig:cnn}
\end{figure}


We train the data using the 127x127 pixel size images and $g,r,i,z,y$ magnitudes as the inputs and the spectroscopic redshift as the label. We use \ntrain galaxies for training, \nval galaxies for validation, and \ntest galaxies for testing. The validation set is used for hyperparamter optimization while the testing data set is used for final performance evaluation since that dataset has been isolated from the model. We trained the model for 200 epochs and utilized a batch size of 256. During training, we utilize the custom loss function from \cite{tanaka2018g}:

\begin{equation}\label{eq:loss}
    L = 1 - \frac{1}{1 + (\frac{ z_{phot}-z_{spec}}{0.15(1 + z_{phot})})^2},
\end{equation}
where $z_{phot}$ is the photometric redshift prediction and $z_{spec}$ is the spectroscopic redshift.

We also use the Adam optimization algorithm with a learning rate of 0.0001 (\cite{kingma2014}).
The training was performed on a desktop computer  using an AMD Ryzen Threadripper PRO 3955WX with 16-Cores and NVIDIA RTX A6000 GPU with 48 GB of GPU RAM. The training used the $127\times127$ pixel dataset and 5-band photometry typically took 48-96 hours and used 20 GB of GPU memory.


To evaluate the performance of the two types of neural networks, we use common metrics for photometric redshift estimates: bias, scatter, RMS error, and outlier rate. The bias, $b$, is defined as:
\begin{equation}
b = {{ z_{phot}-z_{spec} } \over {1+z_{spec}}}.
\end{equation}
The RMS error is:
\begin{equation}
rms =  \sqrt { {{1} \over {n_{gals}}} \Sigma_{gals} \left( {{ z_{phot}-z_{spec} } \over {1+z_{spec}}} \right) ^2 },
\end{equation}
where ${n_{gals}}$ is the number of galaxies. 
The scatter or dispersion is defined with respect to the median absolute error (MAD):
\begin{equation}
    \sigma_{scatter} = 1.48 \times \mathrm{Median}(|\Delta z_i - \mathrm{Median}(\Delta z_i)),
\end{equation}
where $\Delta z_i =  z_{phot,i}-z_{spec,i}$ for the $i$th galaxy. Note the scale factor of 1.48 is to make the MAD more comparable to the standard deviation of a normal distribution.
The outlier, $O$, is defined as:
\begin{equation}
O = {{\vert z_{phot}-z_{spec} \vert} \over {1+z_{spec}}} > .15.
\end{equation}
The outlier fraction is defined as the fraction of outliers divided by the total number of galaxies:
\begin{equation}
    f_{outlier} = N_{outliers}/N_{total}.
\end{equation}
We also use the mean-squared error and the loss as additional metrics. We evaluate these performance metrics in different redshift bins to study the model performance at different redshifts. The number of galaxies in the training set varies significantly between different redshift bins so it is important to assess potential biases with redshift.

We find that the CNN models using images show better predictive performance than the NN models using only photometry (Table \ref{tab:metrics}). The biggest improvement is the bias where the CNN outperforms the NN by about a factor of 10 on average (NN $b = 0.0011$, CNN $b =0.00010$). In addition, the improvement in the model performance with the CNN compared to the NN is larger at higher redshifts. This gap in performance is especially large at the $z=1.5$, the highest redshift we consider in this example. The number of galaxies in the training set at this redshift bin is significantly smaller than at other redshift bins. This likely indicates that the images can encode more information about the redshift than photometry alone, which allows the CNN to achieve higher performance with a smaller training sample compared to the NN with photometry.

\begin{table}[]
    \centering
    \begin{tabular}{lccccccccc}
    \hline\hline
Model &  N galaxies & Loss & Bias &  Scatter &  Outlier Fraction &  MSE \\ \hline
CNN &  40914 & 0.0577 &  0.000100 &   0.0167 &       0.0377 & 0.0615 \\
NN &  40914 & 0.106   & 0.00110   &  0.0315 &        0.0688 & 0.133 \\    \hline
    \end{tabular}
    \caption{Overall performance metrics for NN and CNN models}
    \label{tab:metrics}
\end{table}

\begin{figure*}[tbh]
 \center
 \includegraphics[width=\linewidth]{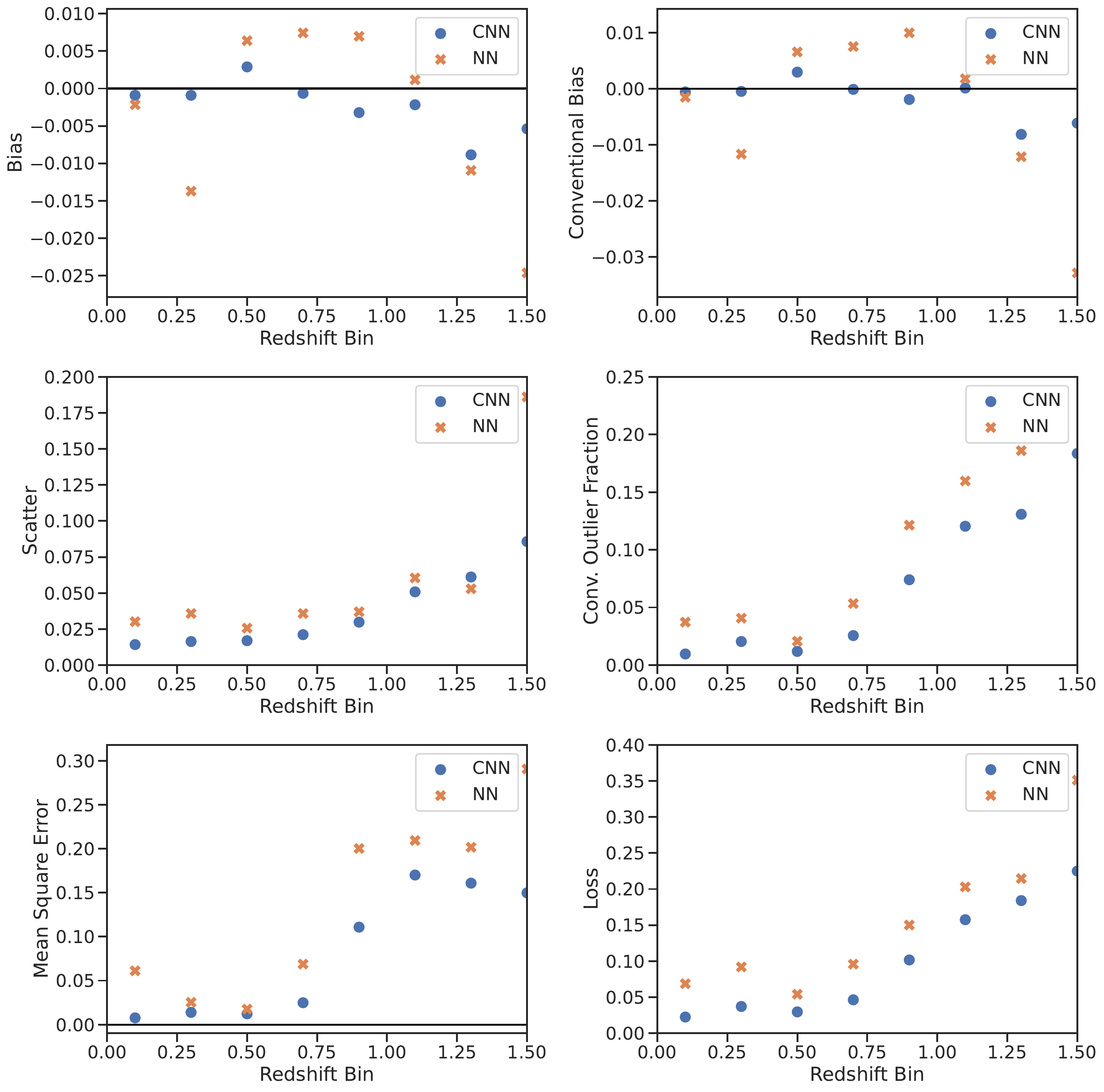}
 \caption{Metrics for the performance of the NN compared to the CNN at different redshift ranges for (a) bias, (b) scatter, (c) outlier fraction, (d) MSE, and (e) loss. The smaller the values are for these metrics, the better the model. The CNN model using images outperforms the NN model using only photometry, especially at larger redshifts where there is less training data.}
 \label{fig:metrics}
\end{figure*}


\end{document}